\documentclass[twocolumn,preprintnumbers,amsmath,amssymb,pra]{revtex4}
\usepackage{amsmath,amssymb,graphicx,epstopdf,bm,soul,upgreek,natbib}
\usepackage{graphicx}
\usepackage{color}
\definecolor{OliveGreen}{rgb}{0,0.6,0}
\usepackage{dcolumn}
\usepackage{bm}
\usepackage{amssymb,float}
\usepackage{cancel}
\usepackage{ulem}
\usepackage{siunitx}
\usepackage[colorlinks,bookmarks=false,citecolor=blue,linkcolor=blue,urlcolor=blue]{hyperref}
\begin{document}
\title{Optically Induced Topological Spin-Valley Hall Effect for Exciton-Polaritons}
\author{R. Banerjee}\email[Corresponding author:~]{rimi001@e.ntu.edu.sg}
\author{S. Mandal}\email[Corresponding author:~]{subhaska001@e.ntu.edu.sg}
\author{T.C.H. Liew}\email[Corresponding author:~]{tchliew@gmail.com}

\affiliation{Division of Physics and Applied Physics, School of Physical and Mathematical Sciences, Nanyang Technological University, Singapore 637371, Singapore}

\begin{abstract}
We consider exciton-polaritons in a honeycomb lattice of micropillars subjected to circularly polarized $(\sigma_\pm$) incoherent pumps, which are arranged to form  two domains in the lattice. We predict that the nonlinear interaction between the polaritons and reservoir excitons gives rise to the topological valley Hall effect, where in each valley two counter propagating helical edge modes appear. Under a resonant pump, $\sigma_\pm$ polaritons propagate in different directions without being reflected around bends. The  polaritons propagating along the interface have extremely high effective lifetime and also show fair robustness against disorder. This paves the way for robust exciton-polariton spin separating and transporting channels, in which polaritons attain and maintain high degree of spin polarization, even in the presence of spin relaxation.
\end{abstract}

\maketitle

{\textit{Introduction.}---} 
Reducing unwanted feedback is one of the key requirements in optical information processing \cite{Science.230.138.1985}. However, when a propagating signal experiences a bend in its path, a significant amount gets reflected. Topological insulators, which are commonly characterized by  gapped bulk modes and  robust edge modes within the bulk band gap \cite{PRL.61.2015.1988}, are often thought of as the potential candidate for transferring signals \cite{NatPhys.7.907.2011,NatNanotech.14.31.2019,Science.365.1163.2019,PRAppl.14.054007.2020,NatPhoton.14.446.2020,Photonics_Research.8.B39.2020}. 

\begin{figure}[t]
\includegraphics[width=0.45\textwidth]{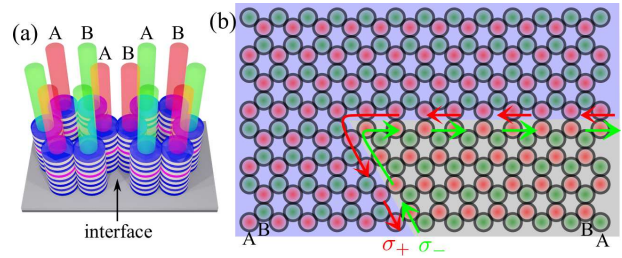}
\caption{Scheme: (a) Circular micropillars arranged in a honeycomb lattice. The two components of the incoherent pump ($\sigma_{\pm}$) are shown in red and green.  (b) Example of two domains that form an interface with a sharp bend. Polaritons with $\sigma_{\pm}$ spins can propagate in opposite directions along the  interface without being reflected. }
\label{Fig1}
\end{figure}

First order as well as higher order topological insulators have been an intense area of research in different fields, such as photonics \cite{PRL.100.013905.2008,NatPhoton.8.821.2014,RevModPhys.91.015006.2019,PRB.98.205147.2018,NatCom.11.3768.2020}, acoustics \cite{PRL.114.114301.2015,NatMater.18.108.2019}, optical lattices \cite{NatPhys.12.639.2016,PRL.123.060402.2019,PRL.122.076801.2019}, etc. The system of exciton-polaritons, where microcavity  photons acquire electronic nonlinearity because of the hybridization with the quantum well excitons, is an excellent platform to study topological phases both in linear \cite{PRX.5.031001.2015,PRB.91.161413.2015,PRL.114.116401.2015,NatPhoton.11.651.2017,Nature.562.552.2018} and nonlinear \cite{Optica.3.1228.2016,LaserPhotonRev.12.1700348.2018,SciRep.7.1780.2017,PRL.119.253904.2017,LaserPhotonRev.13.1900198.2019,PRL.124.063901.2020,OptLett.45.4710.2020,arXiv:2102.13285} regimes. The significant nonlinearity of exciton-polaritons has made a way for different components of an optical information processing device, such as low energy polariton  switches \cite{PRB.90.045307.2014,NatMater.15.1074.2016,OptExp.25.31056.2017}, transistors \cite{OptExp.25.31056.2017,PRB.85.235102.2012,NatCom.4.1778.2013,NatPhoton.13.378.2019}, amplifiers \cite{PRL.109.216404.2012,PRB.93.235301.2016}, memories \cite{PRB.95.235301.2017,NatCom.11.897.2020}, routers \cite{APL.103.201105.2013,PRB.91.195308.2015}, etc. The main motivation in realizing the topological phases is to obtain robust propagation of polaritons which serves to transfer information between the different information processing components \cite{NatCom.10.872.2019}.

The polariton Chern insulator, originally proposed in Refs.~\cite{PRX.5.031001.2015,PRB.91.161413.2015,PRL.114.116401.2015}, is based on the time-reversal symmetry breaking under an applied magnetic field and the transverse electric-transverse magnetic (TE-TM) splitting of the photonic modes, and was realized experimentally in Ref.~\cite{Nature.562.552.2018}. Several other theoretical proposals for realizing topological polaritons followed related to the same scheme \cite{PRB.93.085438.2016,PRB.94.115437.2016,PRB.97.081103R.2018,PRB.98.125115.2018,PRL.122.083902.2019,PRA.99.053836.2019,PRB.100.235444.2019,PRAppl.12.064028.2019,PRAppl.12.054058.2019,OptLett.45.5311.2020}, by using the polarization splitting inside the elliptical micropillars \cite{PRB.98.075412.2018}, by using vortices in staggered honeycomb lattices \cite{NatCom.9.3991.2018}, and by Floquet engineering \cite{PRB.97.195305.2018}. Apart from the linear effects, the nonlinearity of polaritons alone can induce topological phases, such as the appearance of the Haldane model \cite{PRB.93.020502R.2016,PRB.96.115453.2017} and  antichiral edge states \cite{PRB.99.115423.2019}. The non-Hermiticity of the polaritons was used to realize topological phases in 1D micropillar chains \cite{PRR.2.022051.2020,PRL.125.123902.2020,arXiv:2103.05480}.  Even after such advancement of topological polaritonics, the analogue of the topological spin Hall effect \cite{PRL.95.226801.2005}, where different spins propagate in opposite directions, has not been demonstrated yet. This is because the  common schemes for creating topological porlaritons rely on the TE-TM splitting, which is a form of spin relaxation that mixes the two polariton spins corresponding to right and left circular polarizations (denoted $\sigma_{\pm}$). Moreover, the topological bandgap in such cases is  proportional to the TE-TM splitting, which is itself limited.

\begin{figure*}[t]
\includegraphics[width=0.9\textwidth]{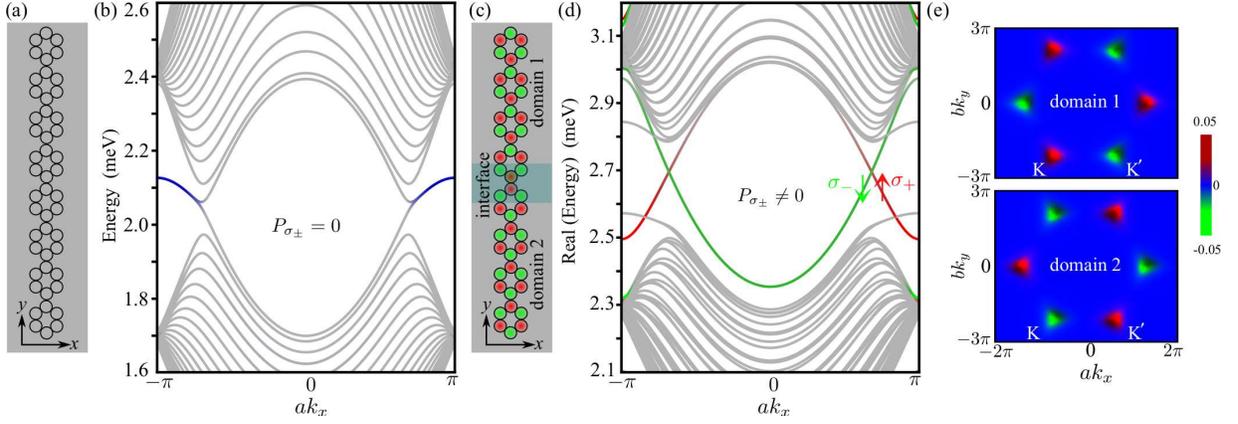}
\caption{(a, c) Honeycomb lattice of micropillars (periodic along the $x$ direction and finite along the $y$ direction) without and with the incoherent pumps, respectively. (b, d) Real part of the band structures of the systems represented in (a) and (c), respectively. The bulk modes are represented in gray. In (b), blue represents topologically trivial modes located at the edges of the sample. In (d), red (green) represents $\sigma_+$ ($\sigma_-$) polarized topological edge states, which are located at the interface. (e) Numerically calculated Berry curvature in the first Brillouin zone for $\sigma_+$ polaritons. The Berry curvature for $\sigma_-$ polaritons is the same as $\sigma_+$ but with the domains interchanged. $b=\sqrt{3}a$ is the periodicity along the $y$ direction. Reservoir parameters used in (c-e): $g_r=10~\mu$eV$\mu$m$^2$, $\tilde{g}_r=-0.4g_r$, $R=3\times10^{-4}$ ps$^{-1}\mu$m$^2$, $\gamma_r=1.5\gamma$, $J=0.09$ ps$^{-1}$, and the peak value of the incoherent pump $P_{\sigma_{\pm}}^{\text{peak}}=10.7$ ps$^{-1}\mu$m$^{-2}$. }
\label{Fig2}
\end{figure*}

There has been a growing interest in realizing  topological phases under the effect of an optical pump \cite{PRL.121.213902.2018,Science.365.1163.2019}. Following a similar route, we consider a honeycomb lattice of circular micropillars, where each micropillar is subjected to a circularly polarized incoherent pump (see Fig.~\ref{Fig1}(a)). The incoherent pumps form two domains in the lattice, where in one domain all the A (B) sublattice sites are subjected to $\sigma_+$ ($\sigma_-$) pumping and vice versa for the other domain. In Fig.~\ref{Fig1} (b), a schematic diagram of two domains is shown. 

We note that lattices of incoherent pump spots are achievable with spatial light modulation techniques \cite{PRL.119.067401.2017,PRL.97.195109.2018,PRL.124.207402.2020,NatCom.11.4431.2020}. In principle a polarizing beam splitter could be used to separate a source laser beam into two oppositely polarized components, which could be modulated differently before being recombined into the required interlocking pattern. Further, in the case of straight interfaces between domains, one polarization corresponds to a reflected and slightly displaced copy of the other, which would allow the polarizing beam splitter to be applied after the spatial light modulator.

The nonlinear interaction between the polaritons and the reservoir excitons induces valley protected helical edge states at the interface, which is otherwise a topologically trivial system with no bandgap. This is related to the topological valley Hall effect \cite{NJP.18.025012.2016,Nature.578.246.2020}, an analogue of the electronic valley based 2D materials \cite{NatRevMater.1.16055.2016}. Unlike previously studied topological polaritons, here the topology is independent of the TE-TM splitting, which makes the edge modes perfectly spin polarized (in the limit of no TE-TM splitting), and the gain due to the incoherent pump ensures a high effective lifetime (around 200 ps) for the edge modes, or even their condensation (above a threshold). Using full numerical simulations we show that the $\sigma_{\pm}$ polaritons propagate in opposite directions without being reflected even in presence of a sharp bend. This effect is used to realize robust polariton spin channels, where polaritons choose to propagate along a particular channel depending upon their spins. The advantage of the system over topologically trivial polaritonic systems is also evaluated explicitly.

{\textit{The model.}---} The polaritons in the micropillars  can be described by the following driven-dissipative Gross-Pitaevskii equation,
\begin{align}
i\hbar\frac{\partial\psi_{\sigma_{\pm}}}{\partial t}&=\left[-\frac{\hbar^2\nabla^2}{2m}+V(x,y)-i\hbar\frac{\gamma}{2}\right]\psi_{\sigma_{\pm}}+\tilde{g}_rn_{\sigma_{\mp}}\psi_{\sigma_{\pm}}\notag\\
&+\left(g_r+i\hbar\frac{R}{2}\right)n_{\sigma_{\pm}}\psi_{\sigma_{\pm}}+F_{\sigma_{\pm}}(x,y)e^{i\left(k_px-\omega_p t\right)},\label{Eqn1}\\
\frac{\partial n_{\sigma_{\pm}}}{\partial t}&=-\left(\gamma_r+R|\psi_{\sigma_{\pm}}|^2\right)n_{\sigma_{\pm}}+J\left(n_{\sigma_\mp}-n_{\sigma_\pm}\right)\nonumber\\
&+P_{\sigma_{\pm}}(x,y)\label{Eqn2}.
\end{align}
Here $\psi_{\sigma_{\pm}}$ are the wave functions of the polaritons corresponding to the $\sigma_{\pm}$ spins and $n_{\sigma_{\pm}}$ represent the densities of excitons with $\sigma_\pm$ spins in the reservoir. The first term represents the parabolic dispersion of the bare polaritons having mass $m$, which is true near the bottom of the lower polariton branch. $V$ is the potential representing the honeycomb lattice of the micropillars and $\gamma$ is the linear decay rate of the polaritons. $g_r$ ($\tilde{g}_r$) is the nonlinear interaction of the polaritons with the reservoir excitons having same (opposite) spin and $R$ is the condensation rate of the polaritons. $\gamma_r$ is the decay rate of the excitons from the reservoir. $P_{\sigma_{\pm}}$  represents the incoherent pumps (with $\sigma_\pm$ components), which we consider first with a strength fixed below the condensation threshold. $F_{\sigma_{\pm}}$ are the two spin components of a resonant pump which serves to create polaritons with frequency $\omega_p$ and wave vector $k_p$.  The coefficient $J$ represents the spin relaxation of the reservoir \cite{PRL.109.016404.2012}. The terms involving $\tilde{g}_r$ and $J$ are not necessary for our desired effect but are included to be realistic.

\begin{figure}[t]
\includegraphics[width=0.5\textwidth]{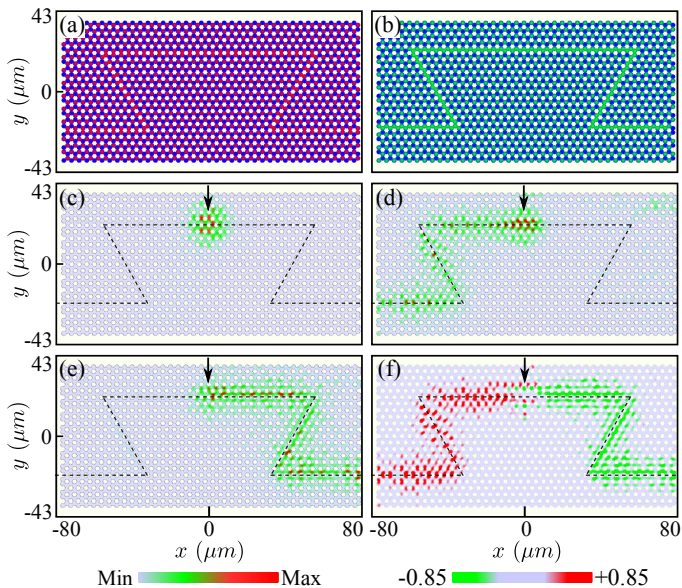}
\caption{(a-b) Arrangement of  $P_{\sigma_\pm}$ in red and green, respectively. (c) Dynamics of the polaritons under a linearly polarized continuous resonant  pump for $P_{\sigma_\pm}=0$. No propagation is observed. The propagations of $\sigma_+$ and $\sigma_-$ polaritons are shown in (d) and (e), respectively, under the same continuous resonant pump as in (c) for $P_{\sigma_\pm}\neq0$. (f) Degree of circular polarization including the TE-TM splitting. The black arrow indicates the position of the continuous resonant pump of width 5 $\mu$m and the dashed line indicates the interface. Parameters: peak value of the resonant pump $F_{\sigma_\pm}^{\text{peak}}=0.01$ meV$\mu$m$^{-1}$, $\omega_p=2.65$ meV/$\hbar$, and $k_p=2\pi/3a$. TE-TM splitting $\Delta_T=50$ $\mu$eV and  $k_T=2.05~\mu$m$^{-1}$ in (f). $P_{\sigma_{\pm}}^{\text{peak}}=7.5$ ps$^{-1}\mu$m$^{-2}$ for the sites subjected to both $P_{\sigma_\pm}$. All other parameters are kept the same as those in Fig.~\ref{Fig2}.}
\label{Fig3}
\end{figure}

We first calculate the linear band structure of the system without the pump and decay by setting $P_{\sigma_{\pm}}=F_{\sigma_{\pm}}=\gamma=\gamma_r=0$. While calculating the band structure, the lattice is considered periodic along the $x$ axis with periodicity $a$ and finite along the $y$ axis (see Fig.~\ref{Fig2}(a)). We choose pillars with diameter 2.5 $\mu$m, potential depth 6.5 meV, $a=4~\mu$m, and $m=3\times 10^{-5}m_e$, where $m_e$ is the free electron mass. The band structure for this case is shown in Fig.~\ref{Fig2}(b), which is similar to that of a Graphene strip with zigzag edges, where the bulk bands (shown in gray) touch at the Dirac points and  trivial edge states (shown in blue) with almost zero group velocity appear. 

Next, we consider the micropillars subjected to the incoherent pumps ($P_{\sigma_{\pm}}\neq0,~ \gamma\neq 0,~\gamma_r\neq0$, but $F_{\sigma_{\pm}}=0$) such that an  interface is formed (see Fig.~\ref{Fig2}(c)). The incoherent pumps create excitons in the reservoir, which interact repulsively with the polaritons and induce a local blueshift. For example, in Fig.~\ref{Fig2}(c) $\sigma_+(\sigma_-)$ polaritons will be blueshifted in the sites shown in red (green). This interaction induced blueshift breaks the inversion symmetry and the bulk bands become gapped. A lattice without any interface (meaning domain 1 or  domain 2 alone) corresponds to a topologically trivial system with gapped bulk but no edge modes within the bandgap \cite{Supp_Mat}. Although, the band structure of both the domains are exactly the same (domain 1 can be transformed into domain 2 by a $180^\circ$ rotation and vice versa), they are topologically distinct. To show this, we calculate the valley projected Chern number for both the domains
\begin{align}\label{Eqn3}
C_{\text{K}}=\frac{1}{2\pi i}\int d^2\mathbf{k}~F(\mathbf{k}),
\end{align}
where $\mathbf{k}=\left(k_x,k_y\right)$, $F(\mathbf{k})=\left(\frac{\partial A_{y}(\mathbf{k})}{\partial k_x}-\frac{\partial A_x(\mathbf{k})}{\partial k_y}\right)$ represents the Berry curvature, $A(\mathbf{k})=\langle u(\mathbf{k})|\nabla_{\mathbf{k}}|u(\mathbf{k})\rangle$ is the Berry connection and  $u(\mathbf{k})$ is the  Bloch mode. Instead of the whole Brillouin zone, the integral in Eq.~(\ref{Eqn3}) is defined around the $\text{K}$ or $\text{K}^\prime$ valley. In Fig.~\ref{Fig2}(e), the numerically calculated Berry curvatures \cite{Supp_Mat} corresponding to the lowest band for $\sigma_+$ polaritons in both the domains are shown. It shows that the Berry curvatures near the K or K$^\prime$ points are opposite in the two domains. The valley projected Chern number turns out to be  $C_{\text{K}(\text{K}^\prime)}=\pm1/2$ in domain 1 and $C_{\text{K}(\text{K}^\prime)}=\mp1/2$ in domain 2. It is easy to see that the difference in valley projected Chern numbers in the two domains is $\Delta C_{\text{K}(\text{K}^\prime)}=\pm1$. The topological bulk-boundary correspondence principle \cite{RevModPhys.82.3045.2010} guarantees the appearance of one edge mode at each valley located at the interface of the two domains and the opposite sign of $\Delta C$ at the two valleys also indicates their counter propagating behaviour. From the symmetry we can argue that $\sigma_+$ at domain 1 and $\sigma_-$ at domain 2 are topologically equivalent, which suggests that the Berry curvature of the $\sigma_-$ polaritons is the same as $\sigma_+$, but with the domains interchanged. This results in $\Delta C_{\text{K}(\text{K}^\prime)}=\mp1$, implying that the $\sigma_-$ edge modes will have opposite group velocity to those of the $\sigma_+$ edge modes. It should be noted that the total Chern number of the system over the whole Brillouin zone is 0. This is why no topological edge mode appears if only one type of domain is considered and it is necessary to form an interface between regions with opposite valley Chern numbers in order to realize the topological edge modes.

We choose the incoherent pumps and reservoir parameters such that the spin dependent blueshift is around 1.5 meV and the degree of circular polarization of the excitonic reservoir is around 17\% \cite{PumpAndResReference}. we take  $\tilde{g}_r=-0.4g_r$, as it is well established that interactions between excitons of opposite spins are attractive in typical cavity polariton systems \cite{PRB.82.075301.2010}. Taking the reservoir into account, the real part of the band structure of the system is presented in Fig.~\ref{Fig2}(d). Indeed at each valley counter propagating $\sigma_{\pm}$ edge modes appear. The band structure calculated for the steady state of the reservoir shows a topological bandgap around 0.3 meV \cite{Supp_Mat}. Being a dynamic system, upon switch on of the incoherent pumping it takes time for the exciton reservoir to build up. The consequence is that the initially trivial system undergoes a topological phase transition in time (this process is illustrated in supplemental movie 1 Ref.~\cite{Supp_Mat}). 
\begin{figure*}[t]
\includegraphics[width=1\textwidth]{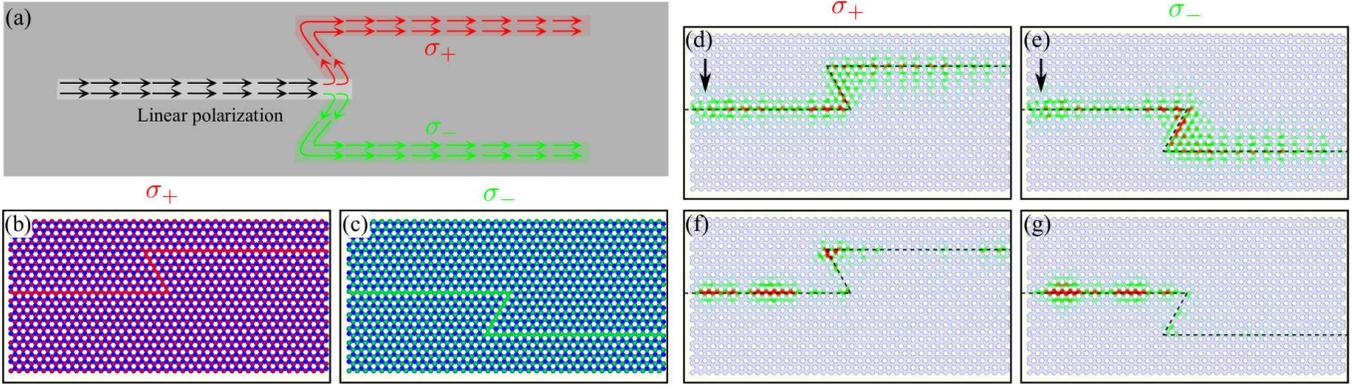}
\caption{(a) Schematic diagram of the polariton spin channels where polaritons choose to  propagate along the upper arm or the lower one depending upon it's spin. (b-c) Arrangement of the incoherent pumps. (d-e) Demonstration of the polariton propagation along the arms depending upon it's spin under a linearly polarized resonant pump. The black arrows indicate the positions of the continuous resonant pump of width 5 $\mu$m. $P^{\text{peak}}_{\sigma_\pm}=7.5$ ps$^{-1}\mu$m$^{-2}$, which is below the condensation threshold. (f-g) Polariton condensates corresponding to $\sigma_\pm$ spins, respectively, for $F_{\sigma_\pm}=0$ and $P^{\text{peak}}_{\sigma_\pm}=10.5$ ps$^{-1}\mu$m$^{-2}$. $\alpha_1=1~\mu$eV$\mu$m$^2$, $\alpha_2=-0.4\alpha_1$. All other parameters are kept the same as those in Fig.~\ref{Fig3}(f).}
\label{Fig4}
\end{figure*}

{\textit{Demonstration of robust polariton transport.}---} Here we consider a honeycomb lattice of micropillars with 40 and 11 unit cells along the $x$ and $y$ direction, respectively. The arrangement of the $\sigma_\pm$ incoherent pumps are shown in Figs.~\ref{Fig3}(a-b), respectively, which forms an interface with sharp bends. We use a Gaussian shaped linearly polarized continuous resonant pump  to inject polaritons in the system.  Fig.~\ref{Fig3}(c) shows the topologically trivial case corresponding to $P_{\sigma_\pm}=0$. Understandably, no propagation of the the polaritons from the excitation spot is observed. Next, the topological case for $P_{\sigma_\pm}\neq0$ is considered. In Figs.~\ref{Fig3}(d-e) the density of the $\sigma_\pm$ polaritons at $t=75$ ps are shown, respectively. As expected, $\sigma_\pm$ polaritons propagate in opposite directions along the  interface with group velocity around 1.9 $\mu$m/ps. The polaritons do not get reflected while propagating around the bend. Since, polaritons with linear polarization get split in space depending upon their spin, this effect can be thought of as an analogue of the topological spin Hall effect \cite{PRL.95.226801.2005}. In general, the propagation of the polaritons is always limited by their short finite lifetime. In this scheme, the incoherent pump ensures a very high effective lifetime of the polaritons (around 200 ps, which is about 6 times larger than the average polariton lifetime considered) \cite{Supp_Mat}. Although we have used resonant pump to inject polaritons, the same can be done by setting the strength of the incoherent pumps $P_{\sigma_\pm}$ above the condensation threshold. In that case, the system spontaneously chooses to condense at the edge modes \cite{Supp_Mat}, similar to the topological insulator lasers.

Unlike the common schemes for creating topological polaritons, here the topological behaviour does not depend upon the TE-TM splitting. This allows us to neglect the TE-TM splitting, which is in principle possible by matching the center of stop band and resonant frequency of the cavity \cite{PRB.59.5082.1999} (which is also the condition sought for the highest quality factor cavities). Nevertheless, to show that the proposed effect is unhampered even in presence of the realistic values of the TE-TM splitting, we add a term $\frac{\Delta_T}{k_T^2} \left(i\frac{\partial}{\partial x} \pm \frac{\partial}{\partial y} \right)^2\psi_{\sigma_\mp}$ at the right hand side of Eq.~(\ref{Eqn1}). The value of the TE-TM splitting $\Delta_T=50~\mu$eV at the wave vector $k_T=2.05~\mu$m$^{-1}$ is taken from Ref.~\cite{PRL.109.036404.2012}. Due to the presence of $\Delta_T$, $\psi_{\sigma_\pm}$ is no longer the eigenstate of the system. Consequently we define the degree of circular polarization as $S_z=\left(|\psi_{\sigma_+}|^2-|\psi_{\sigma_-}|^2\right)/\left(|\psi_{\sigma_+}|^2+|\psi_{\sigma_-}|^2\right)$, which is plotted in Fig.~\ref{Fig3}(f). The robust propagation of the spins in the opposite directions is unaffected (see Ref.~\cite{Supp_Mat}, movie 2), although rather than reaching $\pm1$, the degree of circular polarization is limited to $\pm 0.85$. 

{\textit{Polariton spin channels.}---} Here we show that  rearrangement of the incoherent pumps leads to the realization of   polariton spin channels, where $\sigma_+$ polaritons propagate along the upper arm and $\sigma_-$ polaritons propagate along the lower one. In Fig.~\ref{Fig4}(a), a schematic diagram of such a system is presented. We solve Eqs.~(\ref{Eqn1}-\ref{Eqn2}) in presence of the TE-TM splitting and polariton-polariton interactions (see Ref.~\cite{Supp_Mat}, Eqs.~(S15-S16)) corresponding to the incoherent pump arrangement shown in Figs.~\ref{Fig4}(b-c). The system works as  spin channels under a linearly polarized  continuous resonant excitation (see Figs.~\ref{Fig4}(d-e)) as well as for an incoherent excitation above the condensation threshold, where the condensate forms at the topological edge mode (see Figs.~\ref{Fig4}(f-g)). It can also be noted that the topologically trivial channels show no separation of spins. In principle, one could rely on the optical spin Hall effect \cite{PRL.95.136601.2005} to separate spins in channels \cite{PRB.91.075305.2015}, however, this results in multiple oscillations of the spin.

%{\color{blue}Fig.~\ref{Fig4}, suggests that more than one
%topologically protected interface states can be induced
%throughout the bulk by proper choice of the noresonant pump's profile unlike conventional topological
%insulators, where topological protected states appear at the edges of the physical sample and bulk remains completely unused. In this way, information can be
%transferred throughout the bulk of the sample instead of
%the edges only, making it more compact and scalable.}

{\textit{Discussion and conclusion.}---} 
In conventional photonic topological systems, where topology is induced by (effective) magnetic field (or complex hopping), the topological protection is at the edges of the physical sample, whereas the bulk of the sample remains completely unused. This limits the compactness of the device. However, in our  scheme this is not the case; as the topology is induced optically, more than one topologically protected reconfigurable interface states can be induced throughout the lattice area . In this way, information can be transferred throughout the lattice area of the sample instead of the edges only, making it more compact.

We have presented a scheme to obtain counter propagating transport of $\sigma_\pm$ polaritons, an analogue of the topological spin Hall effect. In the considered system, the nonlinear interaction of the polaritons and reservoir excitons gives rise to topologically protected helical edge modes at each valley of the honeycomb lattice, which can propagate around a sharp bend without being reflected. The topological behaviour being independent of the TE-TM splitting restricts the mixing of two circular polarizations, which helps to obtain almost pure $\sigma_\pm$ spin propagation even after consideration of realistic value of TE-TM splitting. The presence of the incoherent pumps also ensures a very high effective lifetime of the propagating polaritons. Given its topological nature and fair robustness against disorder \cite{Supp_Mat}, this system can be extremely useful in connecting  spin based polariton devices \cite{PRB.70.035320.2004,NatPhoton.4.361.2010,APL.107.011106.2015} as well as recently realized polariton neural networks \cite{PRAppl.11.064029.2019,NanoLett.20.3506.2020,PRAppl.13.064074.2020}.

{\textit{Acknowledgment.}---} The work was supported by the Ministry of Education, Singapore (Grant No. MOE2019-T2-1-004).

\end{document}

%\makeatletter
%\def\@biblabel#1{[S#1]}
%\makeatother

\title{Supplemental material for \\
Optically Induced Topological Spin-Valley Hall Effect for Exciton-Polaritons}%
\author{R. Banerjee}\email[Corresponding author:~]{rimi001@e.ntu.edu.sg}
\author{S. Mandal}\email[Corresponding author:~]{subhaska001@e.ntu.edu.sg}
\author{T.C.H. Liew}\email[Corresponding author:~]{tchliew@gmail.com}

\affiliation{Division of Physics and Applied Physics, School of Physical and Mathematical Sciences, Nanyang Technological University, Singapore 637371, Singapore}

%\date{\today}

%--------------------------------------------------------------------------------------------------------------------------------------
\maketitle
\section{Band structure without an interface}
\begin{figure}[h!]
\centering
\includegraphics[width=0.8\textwidth]{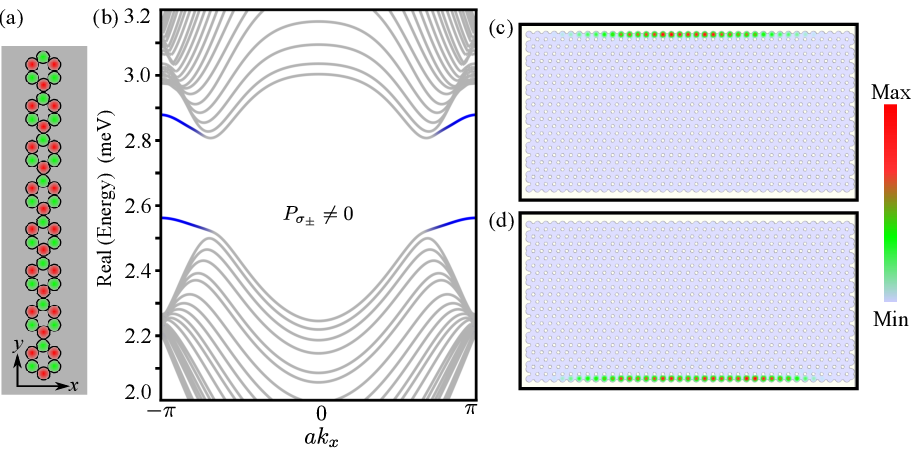}
\caption{(a) A lattice of micropillars subjected to the incoherent pumps $P_{\sigma_\pm}$ without an interface. (b) Band structure of the system. Due to the absence of domain topological edge modes do not appear inside the bandgap. (c-d) Spatial profiles of the modes shown in blue, which are located at the edges of the sample.}
\label{Supp1}
\end{figure}

Here we present the band structure corresponding to a system where the incoherent pumps do not form an interface, i.e. one type of domain only as shown in Fig.~\ref{Supp1}(a). In Fig.~\ref{Supp1}(b) the band structure of such a system is shown. The band structure corresponding to $\sigma_\pm$ polaritons are same. No edge mode inside the bulk bandgap is observed. The almost flat modes shown in blue are located at the edges of the sample (see Fig.~\ref{Supp1}(c-d)).

\section{Calculation of the Valley projected Chern number}
\begin{figure}[t]
\centering
\includegraphics[width=\textwidth]{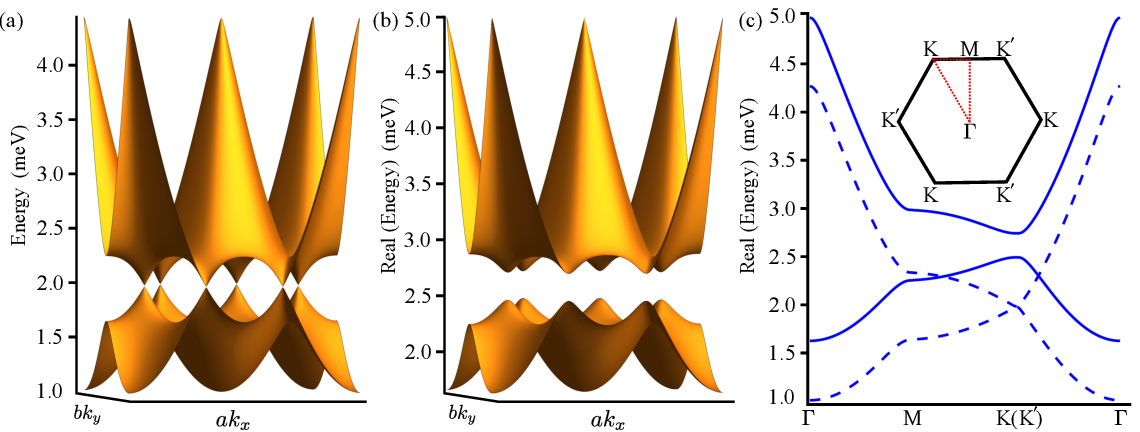}
\caption{(a-b) Two dimensional band structure of the system with no interface for $P_{\sigma_\pm}=0$ and $P_{\sigma_\pm}\neq0$, respectively. The band structures are the same for both the spins in both the domains. (c) The absence and appearance of the bandgap at the K(K$^\prime$)  points for $P_{\sigma_\pm}=0$ (dashed line) and $P_{\sigma_\pm}\neq0$ (solid line), respectively.}
\label{Supp2}
\end{figure}
In this section, we provide the steps for calculating the valley projected Chern number used in the main text. Let us rewrite Eq.~(1) in the main text in the limit of the steady state of the reservoirs and $F_{\sigma_\pm}=0$,
\begin{align}\label{SE1}
i\hbar\frac{\partial \psi_{\sigma_\pm}}{\partial t}=\left[-\frac{\hbar^2\nabla^2}{2m}+V(x,y)-i\hbar\frac{\gamma}{2}\right] \psi_{\sigma_\pm}+\tilde{g}_r n^s_{\sigma_\mp} \psi_{\sigma_\pm}+\left(g_r+i\hbar\frac{R}{2}\right)n^s_{\sigma_\pm} \psi_{\sigma_\pm}.
\end{align}
Here $n^s_{\sigma_\pm}$ represent the steady state of the reservoirs such that $\partial n^s_{\sigma_\pm}/\partial t =0$. In what follows, we'll consider only the Hermitian part of Eq.~(\ref{SE1}) for $\sigma_+$ polaritons by dropping the gain and decay terms and considering the potential $V$ periodic in both the $x$ and the $y$ directions. The modified form of Eq.~(\ref{SE1}) can be written as
\begin{align}\label{SE2}
i\hbar\frac{\partial \psi}{\partial t}=\left[-\frac{\hbar^2\nabla^2}{2m}+V_{\text{eff}}(x,y)\right]\psi,
\end{align}
where $V_{\text{eff}}=V+\tilde{g}_rn^s_{\sigma_-}+g_rn^s_{\sigma_+} $ is the effective 2D periodic potential, which corresponds to $P_{\sigma_+}$ without an interface. Next, we apply the Bloch theory on the Eq.~(\ref{SE2}):
\begin{align}
\psi(x,y)&=u_{k_x,k_y}(x,y)e^{i\left(k_xx+k_yy\right)},\label{SE3}\\
\text{where,  } u_{k_x,k_y}(x,y)&=u_{k_x,k_y}(x+a,y+b),\\
\text{and } V_{\text{eff}}(x,y)&=V_{\text{eff}}(x+a,y+b).
\end{align}
Since, $u_{k_x,k_y}(x,y)$ and $V_{\text{eff}}(x,y)$ are periodic, they can be written as
\begin{align} 
u_{k_x,k_y}(x,y)&=\frac{1}{ab}\sum_{G_x,G_y}\tilde{u}_{k_x,k_y}(G_x,G_y)e^{i\left(G_xx+G_yy\right)},\\
\text{and, }V_{\text{eff}}(x,y)&=\frac{1}{ab}\sum_{G_x,G_y}\tilde{V}_{\text{eff}}(G_x,G_y)e^{i\left(G_xx+G_yy\right)}.
\end{align}
Here, $(G_x, G_y)$ is the reciprocal lattice vector and
\begin{align}
\tilde{V}_{\text{eff}}(G_x,G_y)=\int_{\text{unit cell}}V_{\text{eff}}(x,y)e^{-i\left(G_xx+G_yy\right)}~dxdy\label{SE8}.
\end{align}
Substituting Eqs.~(\ref{SE3}-\ref{SE8}) in Eq.~(\ref{SE2}) we get the following eigenvalue equation, which can be diagonalized numerically to obtain the 2D band structure and Bloch modes:
\begin{align}\label{SE9}
-\frac{\hbar^2}{2m}\left[\left(k_x+G_x\right)^2+\left(k_y+G_y\right)^2\right]\tilde{u}_{k_x,k_y}(G_x,G_y)+\frac{1}{ab}\sum_{G^{\prime}_x,G^{\prime}_y}\tilde{V}_{\text{eff}}(G_x-G^{\prime}_x,G_y-G^{\prime}_y)\tilde{u}_{k_x,k_y}(G^{\prime}_x,G^{\prime}_y)=E\tilde{u}_{k_x,k_y}(G_x,G_y).
\end{align}
Following similar steps, the eigenvalue equation corresponding to $\sigma_-$ polaritons can also be obtained. In Figs.~\ref{Supp2}(a-b) the 2D band structure corresponding to $P_{\sigma_\pm}=0$ and $P_{\sigma_\pm}\neq0$ is shown, respectively. Fig.~\ref{Supp2}(c) shows the difference between the two band structures. It should be noted that the band structures for both $\sigma_\pm$ spins  corresponding to both the domains are the same.

The expression for the Chern number provided in Eq.~(3) of the main text is for the case where the Brillouin zone (BZ) is continuous. Instead, here we deal with a discrete BZ. One obvious choice for calculating the Chern number is to substitute the derivatives and integrals with discrete differences and summation, respectively. However, this procedure of straight forward substitutions often leads to convergence issues, especially for the case where the space is modelled as continuous (as opposed to a tight binding model). To avoid this problem, we define the following quantities relating to the $U(1)$ Gauge variable \cite{JPhysSocJpn.74.1674.2005}:
\begin{align}
U_x(k_x,k_y)=\frac{\int_{\text{unit cell}}u^*_{k_x,k_y}(x,y)u_{k_x+\Delta k,k_y}(x,y)~dxdy}{\left|\int_{\text{unit cell}}u^*_{k_x,k_y}(x,y)u_{k_x+\Delta k,k_y}(x,y)~dxdy\right|},\\
U_y(k_x,k_y)=\frac{\int_{\text{unit cell}}u^*_{k_x,k_y}(x,y)u_{k_x,k_y+\Delta k}(x,y)~dxdy}{\left|\int_{\text{unit cell}}u^*_{k_x,k_y}(x,y)u_{k_x,k_y+\Delta k}(x,y)~dxdy\right|}.
\end{align}
Here, $\Delta k$ is the grid point spacing in the reciprocal space along both the $x$ and $y$ directions. The integration over one unit cell is performed numerically. The Berry curvature is defined as 
\begin{align}
F(k_x,k_y)=\frac{1}{i}\ln\left[\frac{U_x(k_x,k_y)U_y(k_x+\Delta k,k_y)}{U_x(k_x,k_y+\Delta k)U_y(k_x,k_y)}\right].
\end{align}
The above quantity is plotted in Fig.~2(e) in the main text for both domains. The valley projected Chern number can be obtained by summing $F(k_x,k_y)$ near the valleys or over the half BZ:
\begin{align}
C_\text{K}=\frac{1}{2\pi}\sum_{k_x,k_y}F(k_x,k_y).
\end{align}

\section{Effective lifetime of the edge modes}
\begin{figure}[H]
\centering
\includegraphics[width=0.8\textwidth]{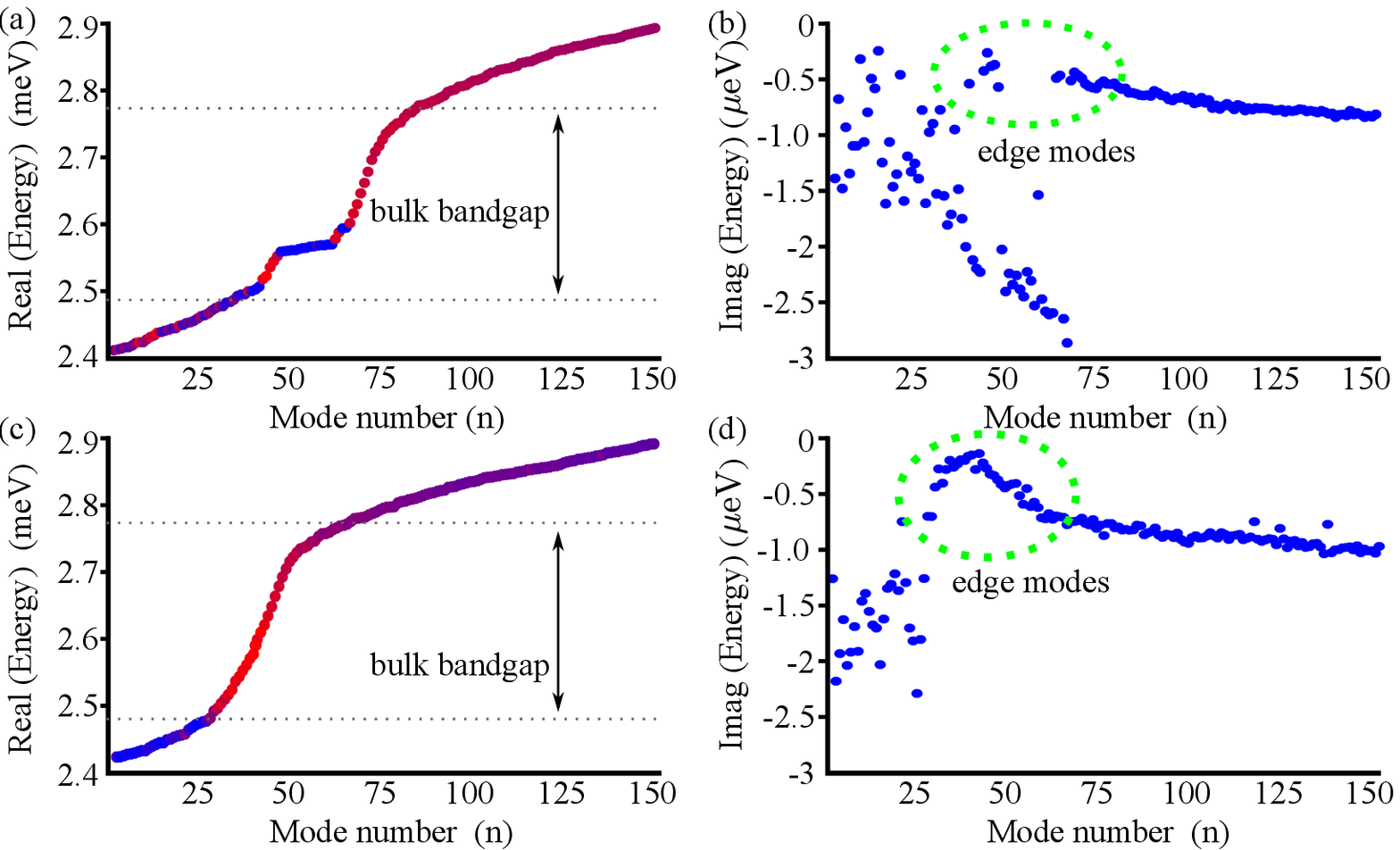}
\caption{ Real and imaginary eigenenergies of the system considered in Fig.~3 in the main text. The real parts of the energies of the $\sigma_\pm$ polaritons in (a, c), respectively, are colour coded corresponding to their imaginary parts in (b, d) with red being the states with less decay. The green dashed circle in (b, d) corresponds to the edge modes.}
\label{Supp4}
\end{figure}

In order to obtain the effective lifetime of the edge modes, we diagonalize Eq.~(\ref{SE1}) corresponding to the incoherent pump arrangement shown in Fig.~3 (a-b) in the main text. The real and imaginary parts of the eigenvalues near the topological band gap are plotted in Fig.~\ref{Supp4}. The effective lifetime of the modes is defined as,
\begin{align}
\tau^n_e=-\frac{\hbar}{2}\times\frac{1}{\text{Imag}\left(\text{Energy}\right)},
\end{align}
where `Real' and `Imag' represent the real and imaginary parts of the Eigen energy, respectively.

\section{Polariton topological insulator laser}
\begin{figure}[H]
\centering
\includegraphics[width=0.8\textwidth]{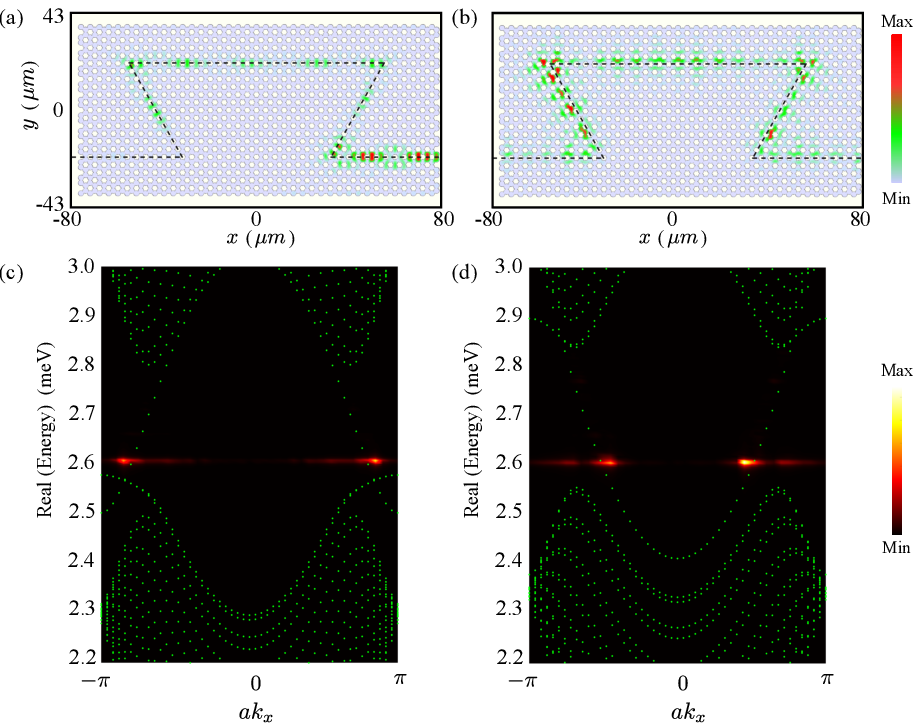}
\caption{ Formation of polariton condensate for $\sigma_+$ spin in (a, c) and for $\sigma_-$ spin in (b, d). Parameters: $F_{\sigma_\pm}$=0, $P^{\text{peak}}_{\sigma_\pm}=13.5$ ps$^{-1}\mu$m$^2$. All other parameters are kept the same as those in Fig.~3(f) of the main text. }
\label{Supp5}
\end{figure}
Here we set the strength of the incoherent pumps above the condensation threshold. In this regime, the nonlinear polariton-polariton interactions become significant and can no longer be neglected. Consequently, we consider the nonlinear  driven-dissipative Gross-Pitaevskii equation,
\begin{align}
i\hbar\frac{\partial\psi_{\sigma_{\pm}}}{\partial t}&=\left[-\frac{\hbar^2\nabla^2}{2m}+V(x,y)-i\hbar\frac{\gamma}{2}\right]\psi_{\sigma_{\pm}}+\left(\alpha_1\left|\psi_{\sigma_{\pm}}\right|^2+\alpha_2\left|\psi_{\sigma_{\mp}}\right|^2\right)\psi_{\sigma_{\pm}}\notag\\
&+\frac{\Delta_T}{k_T^2} \left(i\frac{\partial}{\partial x} \pm \frac{\partial}{\partial y} \right)^2\psi_{\sigma_\mp}+\tilde{g}_rn_{\sigma_\mp}\psi_{\sigma_\pm}+\left(g_r+i\hbar\frac{R}{2}\right)n_{\sigma_{\pm}}\psi_{\sigma_{\pm}}+F_{\sigma_{\pm}}(x,y)e^{i\left(k_px-\omega_p t\right)},\label{SEq15}\\
\frac{\partial n_{\sigma_{\pm}}}{\partial t}&=-\left(\gamma_r+R|\psi_{\sigma_{\pm}}|^2\right)n_{\sigma_{\pm}}+J\left(n_{\sigma_\mp}-n_{\sigma_\pm}\right)+P_{\sigma_{\pm}}(x,y)\label{SEq16}.
\end{align}
Here $\alpha_1$ is the  interaction between the polaritons having the same circular polarization, whereas $\alpha_2$ is the interaction between the polaritons having opposite circular polarization. We take $\alpha_1=1~\mu$eV$\mu$m$^2$ \cite{PRB.100.035306.2019}. We also fix $\alpha_2=-0.4\alpha_1$, which is consistent with experiments, where $\alpha_2$ was experimentally shown to be in the range between 0 and $-\alpha_1$  depending upon  exciton-photon detuning \cite{PRB.82.075301.2010}. Next, we fix $F_{\sigma_{\pm}}=0$ and increase $P_{\sigma_{\pm}}$. Once $P_{\sigma_{\pm}}$ surpasses the condensation threshold, polariton condensation at the topological edge modes for both the spins is observed. This is similar to the topological insulator laser, which is widely studied in photonics \cite{Science.359.eaar4003.2018,Science.359.eaar4005.2018,Nature.578.246.2020,LaserPhotonRev.14.2000001.2020} as well as in polaritonics \cite{PRL.122.083902.2019}. In Fig.~\ref{Supp5} the  profiles of the condensate in the real and the reciprocal space are shown.

\section{Advantage over topologically trivial polariton systems} 
\begin{figure}[H]
\centering
\includegraphics[width=0.7\textwidth]{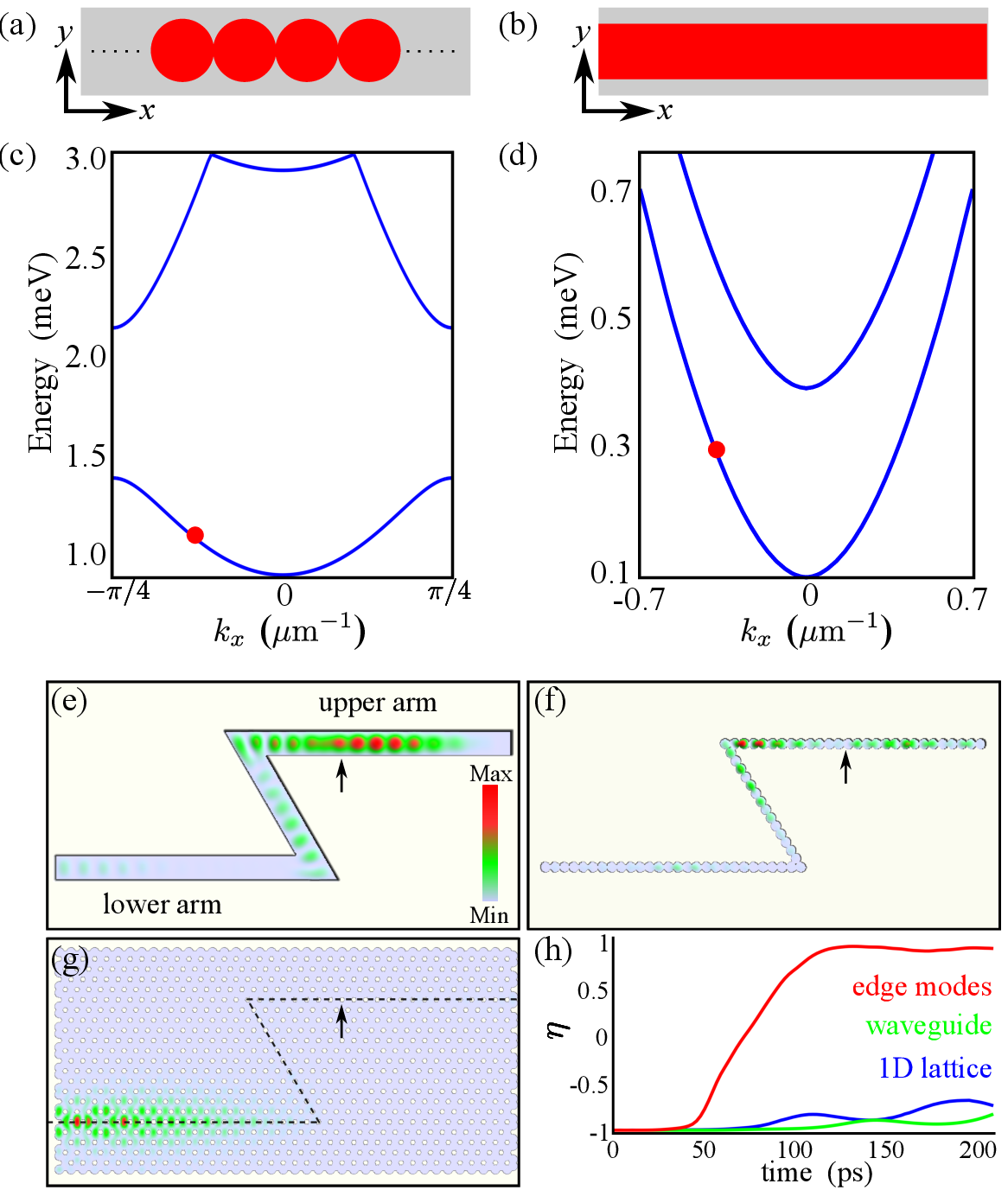}
\caption{ (a-b) Periodic 1D lattice of micropillars and waveguide, respectively. (c-d) The dispersions of the systems shown in (a-b), respectively. The group velocity of the systems is similar to the topological edge modes near the red dot shown in (c-d). (e-g) A snapshot of polariton propagation under a resonant pulse at $t=180$ ps in a 1D waveguide channel, in a 1D lattice of micropillar, and in the proposed topological system (for $\sigma_+$ polaritons only), respectively. The black arrow indicates the position of the Gaussian resonant pulse having width $10~\mu$m and duration 35 ps. (h) Comparison of the efficiency  (defined in Eq.~(\ref{SEq17})) of the three systems. Reservoir parameters used in (g): $g_r=10~\mu$eV$\mu$m$^2$, $R=10^{-3}$ ps$^{-1}\mu$m$^2$, $\gamma_r=1.5\gamma$, and the peak value of the incoherent pump $P_{\sigma_{\pm}}^{\text{peak}}=2.3$ ps$^{-1}\mu$m$^{-2}$. The energy and wave vector of the resonant pulse used in (g) are 1.2 meV and 2$\pi$/15 $\mu$m$^{-1}$, respectively.}
\label{SuppMain4}
\end{figure}
To show the advantage of the proposed system over topologically trivial polariton systems, we additionally consider a 1D waveguide channel and a 1D lattice of micropillars. The shape of the waveguide and the 1D lattice are kept the same as the interface created by the incoherent pumps in the honeycomb lattice (see Figs.~\ref{SuppMain4}(e-g)). Next, we launch a resonant pulse at the upper arm and record the  intensities in the upper ($I_u$) and lower arm ($I_l$) as a function of time. The efficiency of the systems is defined as, 
\begin{align}
\eta=\frac{I_l-I_u}{I_l+I_u}\label{SEq17}. 
\end{align}

For a fair comparison, the average lifetime and group velocity of the polaritons in the waveguide and 1D lattice are matched with those of the edge states. For the topological system, we use micropillars having diameter 4 $\mu$m, lattice periodicity 5 $\mu$m and incoherent pump spots having FWHM around 2.3 $\mu$m. We use only $\sigma_+$ incoherent pumps and set the reservoir parameters corresponding to a blueshift around 0.5 meV. To show the advantage of the topological modes, we work with only one type of polariton spins ($\sigma_+$), and consequently we ignore the spin relaxation of the reservoir. For 1D lattice, the micropillars have diameter $4~\mu$m with periodicity $4~\mu$m. The width of the waveguide is taken as $10~\mu$m. The potential depth and mass of the polaritons are taken as $6.5$ meV and $3\times 10^{-5}m_e$, respectively, in both the cases. In Figs.~\ref{SuppMain4}(a-d) schematic diagrams of the systems and their corresponding bandstructures are shown. We excite  both systems corresponding to the red dots as shown in Figs.~\ref{SuppMain4}(c-d), which have similar group velocity to the edge modes.

In Fig.~\ref{SuppMain4}(h) $\eta$ is plotted as a function of time for the three systems. Almost 100$\%$ transmission of the polaritons from the upper arm to the lower one is obtained for the topological case (see movie 3). On the other hand, because of the significant backscattering around the sharp bends in the waveguide and in the 1D lattice, $\eta$ is negative, indicating that most of the polaritons created by the resonant pulse at the upper arm can not reach the lower arm. This is also consistent with recently studied coupled waveguide systems \cite{AdvOptMater.8.2000650.2020}.

\section{Robustness against disorder}
\begin{figure}[H]
\centering
\includegraphics[width=0.5\textwidth]{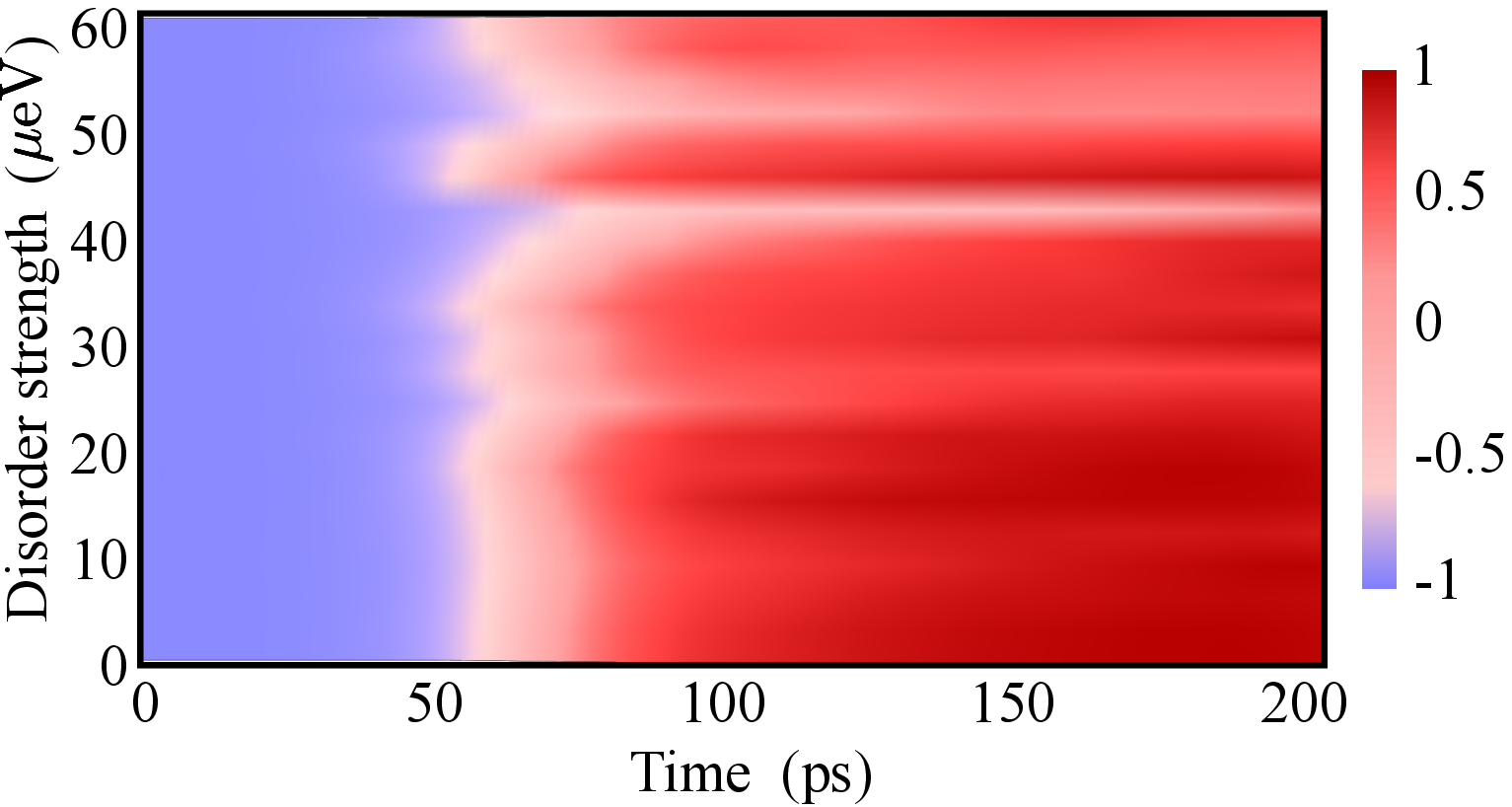}
\caption{ The efficiency $\eta$ as a function of time and disorder strength, which is given by the root mean square value of the random disorder potential.}
\label{Supp7}
\end{figure}

Disorder is always present in practice. Consequently, we add a disorder potential $V_{\text{dis}}$ with the honeycomb lattice potential  $V$. $V_{\text{dis}}$ is taken as a random Gaussian correlated disorder potential having correlation length 2.5 $\mu$m. We consider the same configuration as the one in Fig.~\ref{SuppMain4}(g) and calculate the efficiency $\eta$ for each disorder realization. In Fig.~\ref{Supp7}, $\eta$ is plotted as a function of time and disorder strength, where for each disorder realization a resonant pulse is launched at the upper arm. The system shows fair robustness against disorders of strength around 40 $\mu$eV, which is higher than the typical disorders (around $20-30~ \mu$eV) present in modern micropillar samples \cite{APLPhotonics.3.116103.2018,PRL.116.066402.2016}.

\pagebreak
\section{Pulse Propagation in a Small Triangular Lattice}
\begin{figure}[H]
\centering
\includegraphics[width=0.8\textwidth]{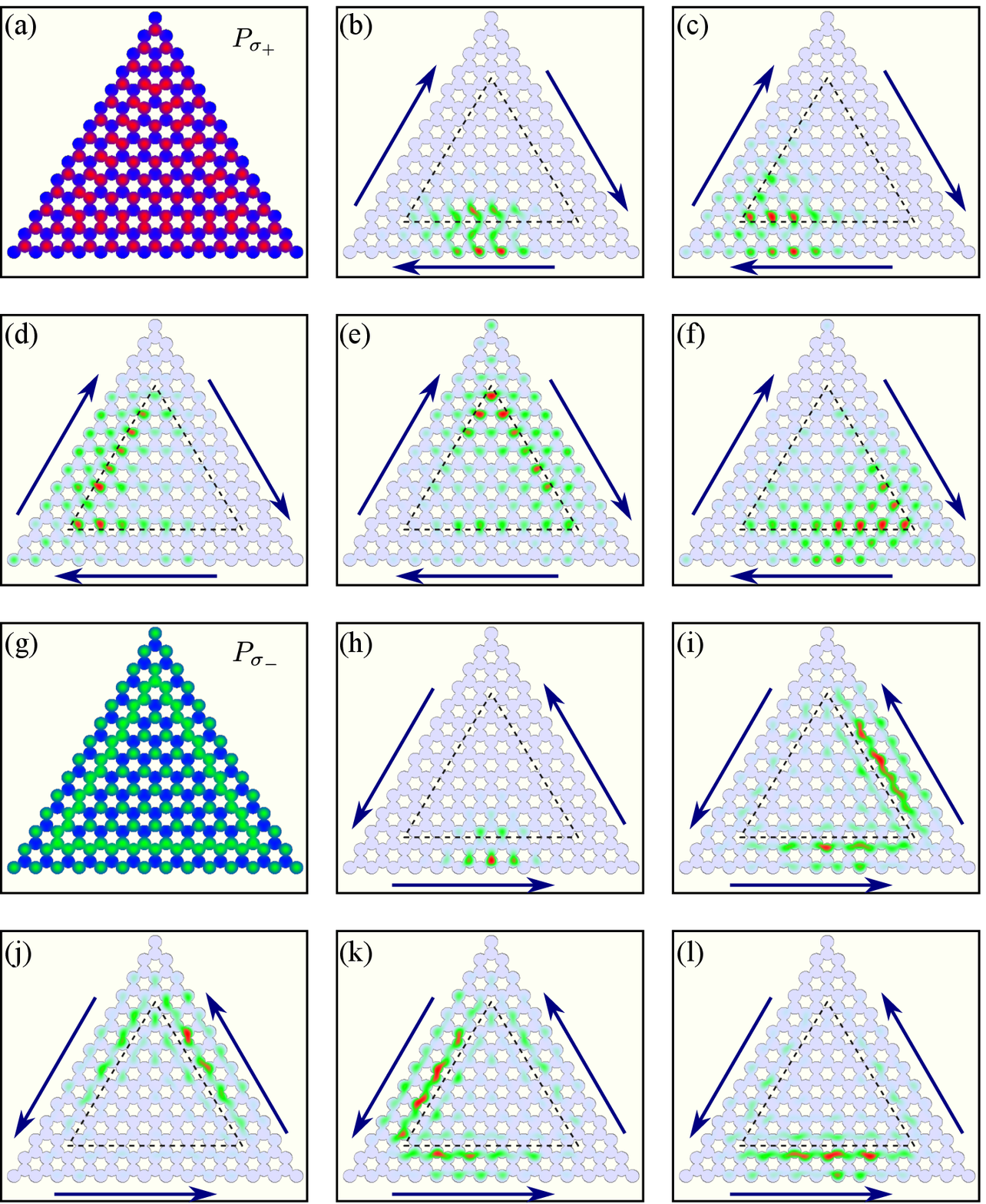}
\caption{ (a), (g) Spatial profiles of $P_{\sigma_\pm}$, respectively. (b-f) Clockwise propagation of $\sigma_+$ polaritons and  (h-l) Counter-clockwise propagation of $\sigma_-$ polaritons under a linearly polarized resonant pulse. Parameters: Pulse width $=10~\mu$m and duration $=35$ ps. All other parameters are kept the same as those in Fig.~3(f) in the main text.}
\label{Supp8}
\end{figure}

For pioneering demonstration, it can be helpful to consider implementation in a small-sized lattice structure. Here, we show that the polaritons having different spins propagate in the opposite directions in a very small triangular lattice geometry having 14 sites in each arm. In Figs.~\ref{Supp8}(a) and (g) the spatial profiles of $P_{\sigma_\pm}$ are shown. Similar hexagonal arrangement of pump spots was shown experimentally in Ref.~\cite{Optica.8.106.2021}. To show the polariton dynamics, we inject the polaritons using a linearly polarized resonant pulse. $\sigma_+$ polaritons propagate in the  clockwise direction (see Figs.~\ref{Supp8}(b-f)), whereas $\sigma_-$ polaritons propagate in the counter-clockwise direction (see Figs.~\ref{Supp8}(h-l)). The full dynamics of the polaritons is shown in Movie 4.

\section{Effect of $\tilde{g}_r$ on the band structure}
In this section, we calculate the band structure using Eqs.~(1-2) in the main text for different values of the $\tilde{g}_r$. As $|\tilde{g}_r|$ decreases the topological bandgap decreases slightly and for $\tilde{g}_r=0$ we get the topological bandgap around 0.25 meV. 

\begin{figure}[H]
\centering
\includegraphics[width=0.7\textwidth]{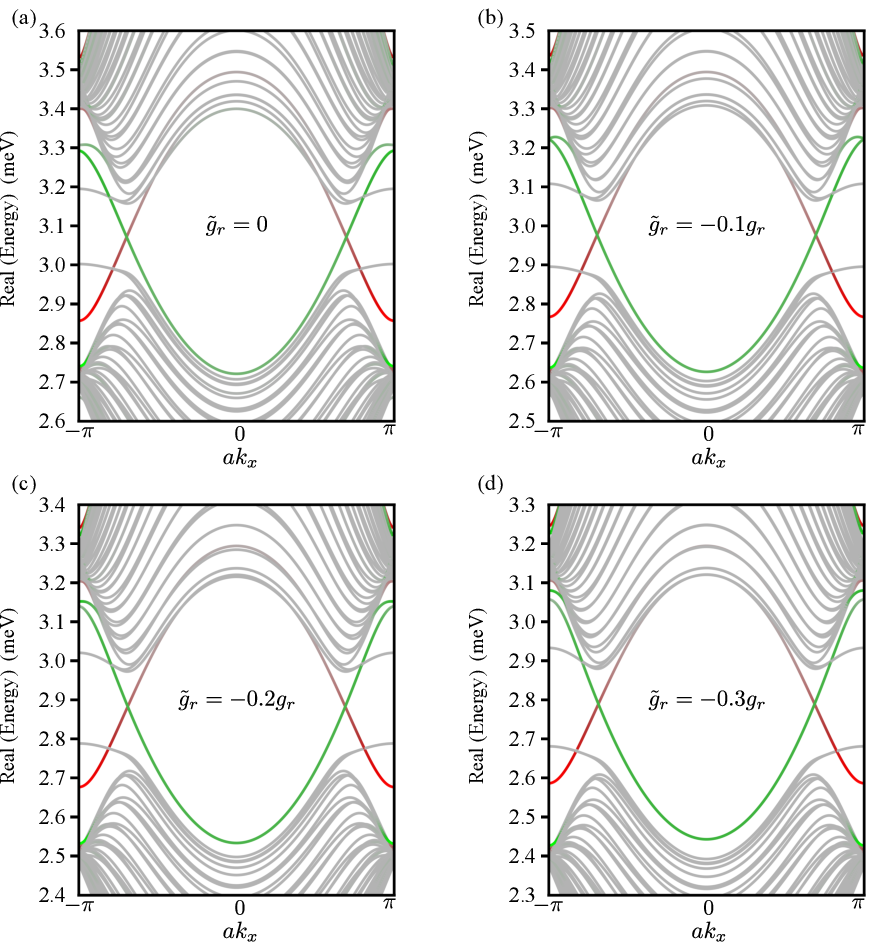}
\caption{(a-d) The dependence of the band structure on $\tilde{g}_r$. All the other parameters are kept the same as those in the main text.}
\label{Supp9}
\end{figure}

\section{Supplementary movies}
{\textit{movie 1.}---} In movie 1, we show the topological phase transition and the appearance of topologically protected edge modes. At $t=0$ a topologically trivial band structure as shown in Fig.~2(b) in the main text is obtained. As the population of the $\sigma_\pm$ excitons increases in the reservoir, a topological band structure with $\sigma_\pm$ edge modes at each valley appear.

{\textit{movie 2.}---} In movie 2, we show that the $\sigma_{\pm}$ polaritons propagate in the opposite directions in the presence of TE-TM splitting. The degree of circular polarization corresponding to $t=250$ ps is plotted in Fig.~3(f) in the main text.

{\textit{movie 3.}---} Movie 3 shows the propagation of the polaritons under the effect of a resonant pulse for the three different systems considered in Figs.~\ref{SuppMain4}(e-g).

{\textit{movie 4.}---} Movie 4 shows the propagation of the polaritons under the effect of a linearly polarized resonant pulse in the triangular lattice geometry shown in Fig.~\ref{Supp8}.